# Exploring the Viability of Unikernels for ARM-powered Edge Computing

Shahidullah Kaiser
*Department of Computer science*
*University of Texas at San Antonio*
San Antonio, TX, USA
shahidullah.kaiser@my.utsa.edu

Ali S¸aman Tosun
*Mathematics and Computer Science*
*University of North Carolina at Pembroke*
Pembroke, NC, USA
ali.tosun@uncp.edu

Turgay Korkmaz
*Department of Computer Science*
*The University of Texas at San Antonio*
San Antonio, TX, USA
Turgay.Korkmaz@utsa.edu

*Abstract*—The rapid expansion of IoT devices and their real-time applications have driven a growing need for edge computing. To meet this need, efficient and secure solutions are required for running such applications on resource-constrained devices with limited power, CPU, and memory. Unikernel, with its minimalistic design and application-specific approach, offers a promising alternative to traditional virtualization and container technologies in these environments. The existing research does not thoroughly examine the feasibility of using unikernel for edge computing. This paper investigates the potential of unikernel for ARM-powered edge computing by evaluating the performance and efficiency of three prominent unikernel systems—OSv, Nanos, and Unikraft—against Docker container. We experiment with real-world edge computing applications and utilize key metrics such as boot time, execution time, memory usage, CPU overhead, and network performance to determine how unikernel performs under the constraints of edge devices. Our findings reveal the potential advantages of unikernel in terms of reduced resource consumption and faster startup times while highlighting areas where they may need further optimization for edge deployment. This study provides valuable insights for researchers and practitioners considering unikernel as a lightweight, efficient solution for edge computing on ARM architectures.

*Index Terms*—Edge Computing, Unikernel, Container, Performance, Osv, Nanos, Unikraft, Docker

## I. INTRODUCTION

The rise of Internet of Things (IoT) devices and their real-time applications has transformed modern computing. These devices generate vast amounts of data and require fast processing, which puts the computational load on traditional cloud computing. The cloud struggles with high latency, network congestion, and security risks. As a result, edge computing has emerged as a critical solution by moving computation closer to the data source. This approach reduces latency, enhances responsiveness, and improves data privacy and security [1]. However, edge computing introduces challenges, especially for devices with limited CPU power, memory, and energy.

ARM-powered devices, commonly used in edge computing due to their energy efficiency and broad adoption, exemplify these constraints [2]. To optimize performance, lightweight and efficient runtime environments are necessary. Containers, such as Docker, are widely used for virtualization in edge environments, but they introduce some overhead and security challenges that can affect performance on resource-constrained devices [3].

Unikernel, a novel virtualization approach, offers a potential alternative to traditional containers. By compiling applications with only the essential parts of the operating system into a single binary, unikernel minimizes resource consumption. This design reduces boot times, lowers memory and CPU usage, and enhances security with a smaller attack surface [4]. However, the practical viability of unikernel in edge environments, especially on ARM architectures, has not been fully explored.

This research explores two critical Edge-to-Cloud Continuum (ECC) components: IoT-to-Edge and Edge-to-Cloud data flow. In the first part, we evaluate whether unikernels can effectively receive and process data from IoT devices at the edge. This involves assessing how well unikernels handle the ingestion and real-time processing of sensor data, considering constraints like low power, limited CPU and memory, and minimal energy consumption. The second part focuses on the Edge-to-Cloud transition, examining whether unikernel can efficiently transmit processed data from the edge to the cloud. This will include testing their ability to manage network communication and data offloading while maintaining low latency and resource efficiency. By addressing these two cases, we aim at determining whether unikernel can serve as a viable solution for end-to-end data processing in edge environments.

The contributions of the paper are as follows:

- This paper explores whether unikernel can be as viable as containers for the Edge-to-Cloud Continuum (ECC). We study the feasibility of Unikraft [5], OSv [6], and Nanos [7] in edge computing. Furthermore, we compare their performance against Docker Container.
- Experiment with the performance of unikernel in edge computing through a data science application designed to receive and process data from nearby devices.
- Investigation of viability of unikernel for running Computer Vision Applications such as Face, Vehicle, Body, and Object Detection applications.
- Identification of metrics for evaluation of unikernel for edge computing scenarios.

This research helps developers assess the potential of unikernel for edge computing by experimenting with new real-world applications in an IoT-Edge environment. It also offers valuable insights for researchers, identifying current trends and opening opportunities for future exploration of unikernel and containerization technologies in IoT and Edge computing. The paper is structured as follows: Section II reviews related work and background, Section III discusses the experimental setup, Section IV presents and analyzes results, and Section V concludes.

## II. BACKGROUND AND RELATED WORK

A Unikernel is a highly specialized, single-purpose system that merges the operating system and application into a compact binary, containing only the essential components required. This streamlined design reduces image size, optimizes resource use, and minimizes the attack surface, which enhances security. Unikernel also offers faster boot times, better resource efficiency, and superior isolation compared to traditional VMs and containers [8]. System calls in unikernel are simple function calls, avoiding costly privilege-level transitions, and context switches are quicker due to the absence of page table and TLB flushes [9]. Their modular design further reduces resource demands like RAM and disk space, showing potentiality for edge computing and IoT.

Unikernel is categorized into two types: language-based and POSIX-based. Language-based unikernel is tightly coupled to specific language runtimes, offering optimized performance but limited flexibility for particular use cases. In contrast, POSIX-based unikernel offers more general compatibility with existing applications while retaining the benefits of single-address space and minimal overhead, making them versatile for edge environments [10]. The broader adoption of language-specific unikernel remains unlikely due to their specialized nature, which limits flexibility and general applicability compared to container-based approaches.

A performance comparison between container, VM, and unikernel was done in [11], [12]. They used a POSIX-based unikernel and concluded that OSv unikernel has approximately half memory consumption compared to VMs and containers. Olivier et al. [9] designed a new unikernel HermiTux and compared the performance against OSv and Rump.

We are not the first one to explore unikernel for edge computing. Mistry et al. [13] introduced prototypical Function-as-a-Service (FaaS) platforms built using language-specific unikernels, IncludeOS (C++), and MirageOS (OCaml). Their findings indicate that unikernel presents a promising alternative to container-based solutions, mainly by reducing cold start times and memory overhead. However, the author did not explore CPU consumption, latency, and implementation of real-world applications to see if unikernel can be used on edge nodes. Besides, Goethals et al. [14] compared OSv unikernel, runc, runsc-based container, and Firecracker with Linux microVMs for Function-as-a-Service (FaaS) deployments. Their findings reveal that OSv unikernel demonstrates extremely low boot times and outperforms Docker containers in terms of memory usage. Moebius et al. [15] explores the potential of unikernel as a secure execution environment for edge FaaS. Through an evaluation of the Linux-compatible Nanos and OSv unikernel toolchains, the authors highlight advantages in cold start efficiency and memory usage.

Existing research has not extensively tested a wide range of unikernels, such as Unikraft, nor has it explored their application in real-world IoT-edge scenarios. In our research, we investigated the readiness of unikernels for edge computing by conducting experiments to evaluate whether unikernels can serve as a viable option in real-time applications such as running data science and computer vision applications.

## III. EXPERIMENTAL SET UP

Edge clusters can be homogeneous (same architecture) or heterogeneous (different architectures). In multitenant and resource-constrained environments, unikernel offers a promising solution due to its lightweight nature and reduced overhead. A critical aspect of edge nodes is handling real-time data from IoT devices. In our experimental setup, we design an IoT-Edge network where edge node receives and processes data from IoT devices. The edge node, represented by a Raspberry Pi 4 (RPi 4), runs a data science application, while a separate machine hosts the IoT device, sending data to the edge via a wireless network. To simulate real-time processing, we utilize Fitbit fitness tracker data [16], specifically the Daily Activity dataset, which includes attributes such as *ActivityDate, TotalSteps, TotalDistance,* and *Calories*.

The edge node's task is to calculate the average number of steps for each user and identify the user with the maximum average steps. Each unikernel and container recorded critical metrics, including processing time, CPU usage, and RAM usage. Processing time, the time taken to perform a specific task on the edge node, was measured by varying the data size in different chunks. We conduct the experiment 10 times, averaging the results for each iteration. This approach allows us to observe the efficiency of each unikernel in resource-constrained environments, highlighting the suitability of unikernel for edge computing. Specifically, Unikraft, Nanos, and OSv are evaluated for their ability to handle real-time IoT data, providing insights into how these unikernels can optimize performance in edge environments. An overview of our experimental set-up is shown in figure 1.

The RPi 4, with its low-power ARM architecture, represents typical edge nodes in these experiments. The combination of real-time IoT data processing and efficient resource usage underlines the practical benefits of unikernel in edge scenarios, where responsiveness and minimal overhead are critical. Table I outlines the specifications of our experimental setup.

For our experiment, we set up RPi 4 with Ubuntu 22.04, a 64-bit operating system. Additionally, we used an ASUS RT-N56U Dual Band Wireless N600 Gigabit router to manage wireless network configurations. This router supports 802.11 N wireless technology and features 10/100/1000 Mbps LAN ports for high-speed connectivity.

In our experiments, we select three POSIX-based unikernel—OSv, Nanos, and Unikraft—because they offer varying degrees of compatibility and optimization for real-world workloads. Their POSIX-like environment allows them to run stan-dard applications with minimal changes, making them suitable for various edge computing scenarios. Besides, orchestration tools are only available for these three unikernels. Using all three, we aim to compare their performance and suitability for different edge workloads.

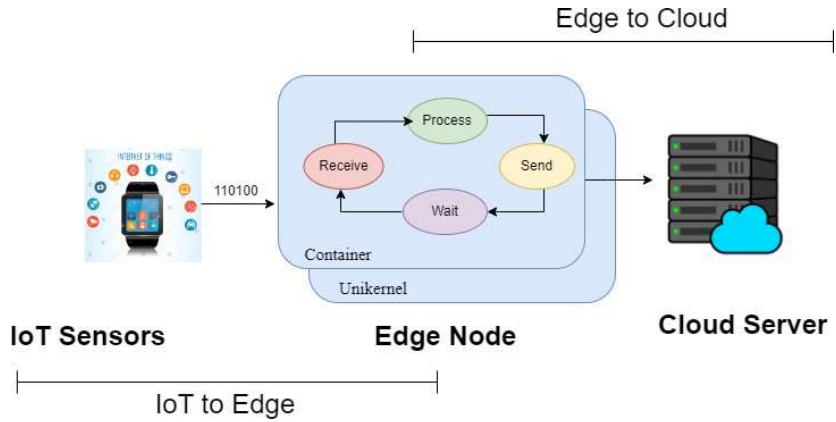

Fig. 1: Overview of our experimental set-up

| Description | Specification |
| --- | --- |
| Architecture | ARM 64 |
| SoC | Broadcom BCM2711 |
| Processor | ARM 1.5GHz CPU |
| Total Cores | Quad-core Cortex A-72 |
| Total RAM | 4GB |
| OS | Linux |
| Power Consumption | 500 mA |
| Micro-SD card | Available |
| Wireless | 2.4 GHz and 5GHz |
| Bluetooth | 5.0 |

TABLE I: Specification of Experimental Setup

## IV. RESULTS AND DISCUSSION

In this section, we present the outcomes of our experiments, comparing the performance of different unikernels with Docker container to assess their suitability for edge computing. We evaluated key metrics such as boot time, processing time, resource usage (CPU and memory), and network latency to gauge the effectiveness of each unikernel in resource-constrained edge environments.

First, we investigate whether a unikernel can serve in an edge node by receiving data from an IoT device. We measure data receiving time across different unikernels and compare them to Docker, varying the data size to assess performance scalability. Then, we test the edge node's ability to forward the received data to the cloud, observing how efficiently unikernel could handle this transfer process compared to traditional containers. The experiment aim to determine whether unikernels, with their lightweight architecture, can meet the demands of real-time data processing in edge-cloud environments.

### A. Boot Time

To investigate how different unikernels start, we measure their boot times using our application. It simulates a real-world IoT-Edge scenario where the unikernel is invoked by receiving data from an IoT device. In edge computing, latency and resource constraints are critical. Besides, faster boot times are essential for maintaining responsiveness, especially when services must be deployed or scaled quickly in dynamic environments. The minimal startup delay would allow edge devices to quickly bring up instances to process incoming IoT data, leading to more efficient real-time data processing. This is crucial for applications like autonomous driving, smart cities, healthcare, and industrial IoT.

Figure 2a illustrates the average boot time for a single instance across Docker, Unikraft, Nanos, and OSv. Docker shows a significantly higher boot time (977.5 ms). In comparison, unikernel like Unikraft (232.25 ms), Nanos (246.94 ms), and OSv (251.75 ms) demonstrate much faster boot times, reducing the delay to nearly one-fourth of Docker's. Besides, running multiple lightweight instances to manage distributed workloads in edge environments is common. The ability to scale efficiently without significant performance degradation is crucial. Unikernels exhibit much better scalability in terms of boot time compared to Docker. Figure 2b shows the boot times of Docker, Unikraft, Nanos, and OSv for increasing numbers of instances. As the number of instances grows, the boot time for Docker scales significantly, reaching over 1400 ms for 30 instances, while the unikernel boot times remain comparatively low and increase only slightly.

The reason for Docker's higher boot time stems from its additional overhead. Operating on top of a complete OS layer, Docker requires more resources and time to initialize components. By contrast, unikernels bundle only the essential elements required for each application, avoiding the excess overhead and thus significantly reducing boot times.

Unikernels, with their reduced boot times, can contribute to minimizing service downtime and ensuring high application availability in resource-constrained edge environments. Since the edge environment often deals with unpredictable network conditions and intermittent connectivity, rapid boot times also support failover and recovery scenarios where new instances must be deployed swiftly in response to node failures.

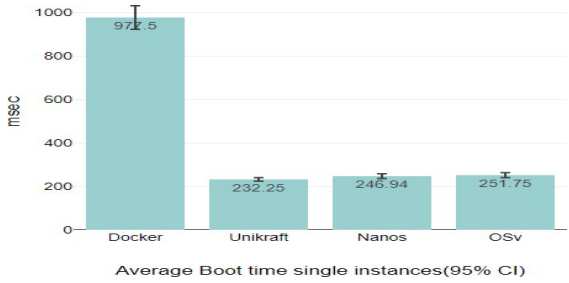

(a) Average Boot Time for Single Instance

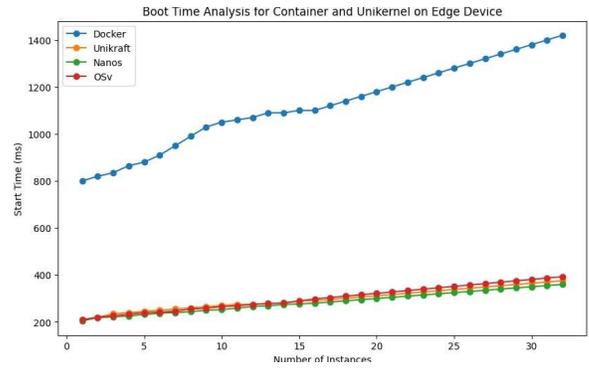

(b) Boot Time for Multiple Instances

Fig. 2: Boot Time Comparison of Container and Unikernel (lower is better)

## B. Processing Time

We investigated how efficiently selected unikernels and container can perform the given task. The evaluation criterion is the processing time: the time taken to perform the given task at the edge. The violin plot in figure 3 provides valuable insights into the processing time performance of Docker, Unikraft, Nanos, and OSv. The task is to calculate the average steps per user and identify the one with the maximum average.

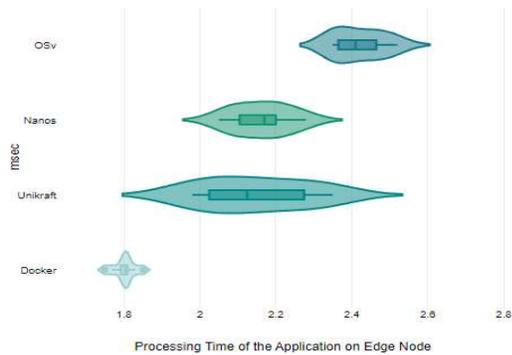

Fig. 3: Processing Time of Container and Unikernel on edge node

Docker demonstrates the lowest processing time, consistently below 2 milliseconds (msec). The compact and narrowest spread indicates Docker's performance is more consistent than the others. This suggests that Docker is highly optimized for handling tasks that require quick response times and faster processing on edge devices. Unikraft shows moderate variability with median value closer to docker. Unikraft is slightly slower than Docker and maintains a competitive performance of around 2 to 2.1 msec. Nanos and OSv exhibit higher processing times, with OSv being the slowest, crossing 2.5 msec for most trials. Besides, OSv has the highest variability in processing time. These results indicate that Docker and Unikraft are better suited to latency-sensitive applications than Nanos and OSv.

## C. CPU Usage

We experimented with how well the CPU resource is utilized on the edge node. Figure 4a illustrates the CPU usage across Docker, Unikraft, Nanos, and OSv when executing the same application on an edge device. Docker demonstrates the highest CPU consumption, with usage ranging between 0.26% to 0.31%, indicating that while Docker offers general-purpose functionality, it incurs significant CPU overhead. OSv shows slightly lower consumption, with values fluctuating from 0.19% to 0.26%, likely due to its more generalized system design, which results in greater resource consumption. Nanos presents a more moderate CPU consumption, spanning between 0.19% and 0.24%, with reduced variability compared to OSv, demonstrating its efficiency while still using more resources than Unikraft. Unikraft showcases the lowest CPU usage, tightly clustered between 0.17% and 0.20%, benefiting from the absence of KPTI (Kernel page-table isolation) and the presence of Mimalloac as a system-wide allocator.

These findings suggest that Unikraft is the most CPU-efficient among the unikernels, making it the optimal choice for CPU-bound tasks in edge computing. Docker, though resource-intensive, offers versatility across various use cases, which can justify its higher CPU utilization in scenarios where flexibility is prioritized. OSv and Nanos offer unique trade-offs, with Nanos being more efficient than OSv but still consuming more resources than Unikraft. Ultimately, the selection between these platforms depends on the balance between CPU efficiency and the functionality required for specific edge computing applications.

## D. Memory Usage

Existing research recommends using simple and lightweight workloads to understand the memory consumption on edge nodes [17]. Our lightweight data science workload meets this requirement. Figure 4b represents the memory consumption of Docker, Unikraft, Nanos, and OSv when running the application on an edge device.

Docker exhibits the highest memory consumption among the platforms, averaging around 60 MB, with slight variance.

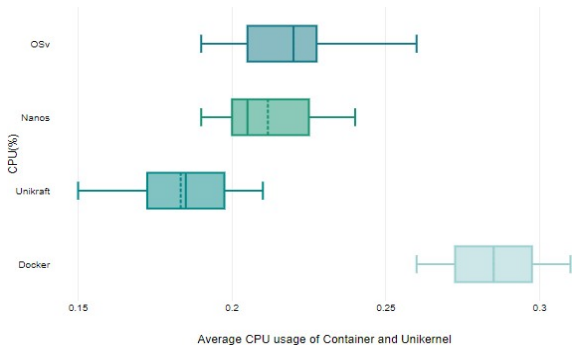
(a) Average CPU usage of Container and Unikernel

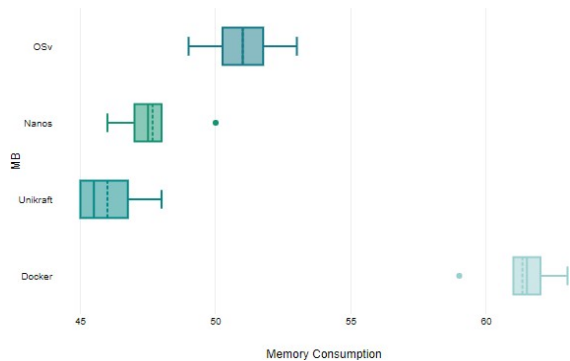
(b) Average memory usage of Container and Unikernel

Fig. 4: Resource Consumption on edge Node

This is expected, as Docker containers usually include an entire operating system layer and libraries, leading to higher overhead. In contrast, Unikraft shows the lowest memory usage, around 45-48 MB, aligning with its goal to build tailored unikernel with minimal components. Nanos also demonstrates low memory usage, slightly higher than Unikraft, but remains efficient, with an average consumption of around 50 MB. Lastly, OSv has higher memory consumption than Unikraft and Nanos, averaging around 55 MB. OSv consumes more memory due to its memory allocation system, particularly how it reserves pages for each vCPU in the L1 and L2 memory pools. Additionally, the way it handles smaller allocations (less than 4096 bytes) can lead to memory waste, as total pages are used even if the request does not fully utilize them.

This comparison highlights the efficiency of unikernels, particularly Unikraft, in reducing memory overhead, which is essential for resource-constrained edge environments. These findings suggest that, for memory-sensitive applications, unikernels like Unikraft or Nanos offer substantial benefits over traditional containers like Docker.

*E. Network Latency*

We investigate how efficiently container and unikernel can handle upcoming data. To observe this, we run an application server in each unikernel and Docker on the edge node. The client sends a request to the server to see how it handles requests from IoT devices. Figure 5 shows the average latency of data requests across Docker, Unikraft, Nanos, and OSv for increasing queries. Docker exhibits a noticeable spike in latency between 200 to 400 requests, peaking at around 14 milliseconds. This behavior is likely due to Docker's higher I/O and process overhead during periods of higher load. As the number of requests increases beyond 500, Docker stabilizes, with latency decreasing and flattening, which could be attributed to its caching mechanism optimizing over time. Meanwhile, Unikraft, Nanos, and OSv maintain relatively consistent, lower latency throughout due to their lightweight architecture and reduced overhead compared to Docker.

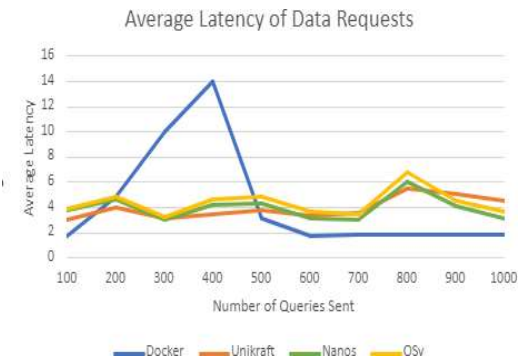
Fig. 5: Number of Queries sent per seconds

We further conducted experiments to measure the time taken by each platform (Docker, Unikraft, Nanos, OSv) to receive data from IoT devices. The receiving time is the interval from when the data is sent to when the complete dataset is received at the edge node. We varied the data size, starting with 91KB (fitbit data) and progressively doubling it up to eight times. Figure 6 shows our experiment on receiving time. As the dataset size increased, network and processing latencies rose proportionally, indicating a consistent relationship between data size and latency across all platforms.

*F. Limitation of Unikernel*

We investigate whether unikernel can be used to perform further operations from the edge node to the cloud. We attempt to send data from the edge node (running a unikernel) to the cloud, but we fail to do the transmission. One possible explanation is the lack of a traditional file system in unikernels. Since they are designed to be minimal and tailored for single applications, they do not typically include robust file-handling mechanisms like those found in full-fledged operating systems. This limitation made accessing and managing the received data for further transmission difficult. Unikraft does offer basic file-handling features and can support simple file systems. Still, these capabilities are limited and may not have been sufficient

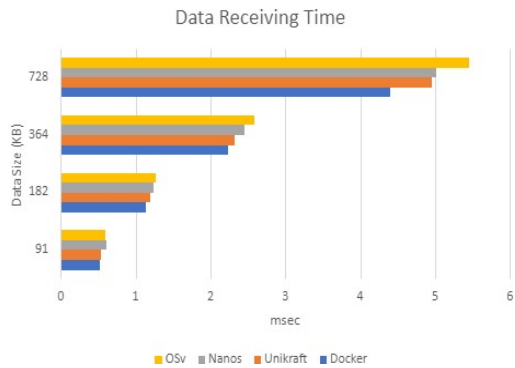

Fig. 6: Time to receive the data from IoT to edge node

for our use case, especially when dealing with complex data transmission to cloud servers.

We also experiment with both container and unikernel to run computer vision applications such as object detection. However, due to the lack of sufficient openCV library support, we failed to do so with unikernel. Fortunately, we run the computer vision application with docker container. We did not include the result as we can not compare it with unikernel. The widespread adoption of unikernel in edge computing presents significant challenges, particularly when applied to computer vision tasks. Unikernels, by design, strip down many traditional OS components to remain lightweight and efficient. But this minimalism prevents them from supporting complex libraries like OpenCV, which are essential for running models such as Haar Cascade, HoG, YOLO, or MobileNet. These models rely on extensive system calls, dynamic memory allocation, and specific hardware drivers that unikernel lacks. Additionally, unikernels are optimized for single-use applications with minimal dependencies, which makes them unsuitable for complex, multi-layered tasks like image recognition and machine learning inference. They also do not support the hardware acceleration, such as GPU access, that computer vision applications often require for real-time processing. Therefore, while unikernels are advantageous in lightweight, resource-constrained environments, their limitations make them a poor fit for high-performance computer vision applications in edge computing. In summary, unikernels are better suited for FaaS (Function-as-a-Service) due to their efficiency, speed, and security, whereas for PaaS (Platform-as-a-Service) they lack the flexibility and ecosystem support.

## V. CONCLUSION

Unikernels offer a smaller image size and minimal memory and CPU consumption compared to containers and VMs. Unikernel is particularly advantageous in scenarios where rapid scaling and resource efficiency are critical. Their lightweight nature enables them to respond quickly to user requests and operate efficiently on resource-constrained edge devices. By minimizing the attack surface, unikernels enhance security, ensuring code integrity and simplifying updates. Their low operating system overhead makes them well-suited for applications requiring frequent context switching while processing small amounts of data. In Edge computing, container, and unikernel technologies will likely coexist, offering flexibility to choose the best-suited virtualization technology for varying performance and resource requirements. In the future, we plan to investigate a hybrid container-Unikernel-based Edge system design. Besides, we plan to build edge clusters of heterogeneous architecture (e.g., X_86 and ARM) and investigate the performance on both architectures. We also plan to investigate to improve the unikernel to avoid limitations so that they can be used for computer vision applications.